# Origin of Increased Curie Temperature in Lithium-Substituted Ferroelectric Niobate Perovskite: Enhancement of the Soft Polar Mode

Hao-Cheng Thong,[1] Fang-Zhou Yao,[2,*] Xian-Xian Cai,[1] Ze Xu,[1] Mao-Hua Zhang,[2] Huazhang Zhang[3], Ben Xu,[4] Yan Wei,[5,*] Shi-Dong Wang[6,*], and Ke Wang[1]

[1] State Key Laboratory of New Ceramics and Fine Processing, School of Materials Science and Engineering, Tsinghua University, Beijing 100084, P. R. China

[2] Research Center for Advanced Functional Ceramics, Wuzhen Laboratory, Jiaxing 314500, P. R. China

[3] School of Physics and Mechanics, Wuhan University of Technology, Wuhan 430070, P. R. China

[4] Graduate School, China Academy of Engineering Physics, Beijing 100193, P. R. China

[5] Department of Geriatric Dentistry; NMPA Key Laboratory for Dental Materials National Engineering Laboratory for Digital and Material Technology of Stomatology, Peking University School and Hospital of Stomatology, Beijing 100081, P. R. China

[6] Musculoskeletal Tumor Center, Peking University People's Hospital, Beijing 100044, P. R. China

* Corresponding author: Fang-Zhou Yao (yaofangzhou@xjtu.edu.cn); Yan Wei (kqweiyan@bjmu.edu.cn); Shi-Dong Wang (bonetumor@bjmu.edu.cn)

**Abstract**

The functionality of ferroelectrics is often constrained by their Curie temperature, above which depolarization occurs. Lithium (Li) is the only experimentally known substitute that can increase the Curie temperature in ferroelectric niobate-based perovskites, yet the mechanism remains unresolved. Here, the unique phenomenon in Li-substituted $KNbO_3$ is investigated using first-principles density functional theory. Theoretical calculations show that Li substitution at the A-site of perovskite introduces compressive chemical pressure, reducing Nb–O hybridization and associated ferroelectric instability. However, the large off-center displacement of the Li cation compensates for this reduction and further enhances the soft polar mode, thereby raising the Curie temperature. In addition, the stability of the tetragonal phase over the orthorhombic phase is predicted upon Li substitution, which reasonably explains the experimental observation of a decreased orthorhombic-to-tetragonal phase transition temperature. Finally, a metastable anti-phase polar state in which the Li cation displaces oppositely to the Nb cation is revealed, which could also contribute to the variation of phase transition temperatures. These findings provide critical insights into the atomic-scale mechanisms governing Curie temperature enhancement in ferroelectrics and pave the way for designing advanced ferroelectric materials with improved thermal stability and functional performance.

# 1. Introduction

Ferroelectric materials are crucial for applications such as actuators, sensors, and memory devices [1-4], each requiring distinct material properties. For instance, a high piezoelectric coefficient is required for actuators and sensors [5-7], while stable, low-field switchable polarization is critical for non-volatile memory devices [8, 9]. However, many of these properties are temperature-sensitive, with the Curie temperature ($T_C$) being particularly important: above $T_C$, ferroelectricity is lost due to depolarization, potentially leading to malfunctions. It is particularly challenging for high-temperature applications, such as sensors or actuators in nuclear reactors and oil drilling [10], where stability of properties at elevated temperatures is critical [11-16]. Consequently, increasing $T_C$ has become a significant research focus in ferroelectrics.

The paraelectric-ferroelectric phase transition is often described by Landau's phenomenological theory and soft-mode theory. Upon cooling down from the $T_C$, the single-well Landau free energy curve becomes a double-well or triple-well curve, depending on whether it is a first-order or second-order phase transition, which is usually referred to as the ferroelectric instability in soft-mode theory [17]. The formation of spontaneous polarization is mainly due to the freezing (known as softening) of the relative displacement of cation and anion (known as optical mode) in ferroelectric perovskite at Γ point. Away from Γ point (*e.g.*, M, R, or X), more complex polar structures may emerge, *e.g.*, antiferroelectric and ferrielectric [18, 19]. Intuitively, combining different cations and anions would result in different paraelectric-ferroelectric phase transition behavior, as the soft polar mode strongly depends on the interaction among these ions [20, 21]. It has been explained based on the competition between the long-range dipole-dipole and short-range interaction [22].

There are few existing experimental methods that can effectively raise $T_C$ in ferroelectrics, with one common strategy being to incorporate a perovskite with an inherently high $T_C$ into the matrix system, forming a solid solution. Li substitution in potassium sodium niobate (KNN) is a prime example of this approach [23]. Demonstrated by Saito *et al.* in 2004 [24]. Li remains the only experimentally known substitute for increasing the $T_C$ of KNN. Machado *et al.* discovered a similar effect in the molecular dynamics simulation of $KNbO_3$ based on the shell model, noting a large off-center displacement of Li as a key factor in raising $T_C$ [25], but without giving a deeper insight into the physical origin. Recently, Li *et al.* re-investigated the enhanced $T_C$ and hardening effect in Li-substituted sodium niobate induced by annealing at slightly below $T_C$ [26], first reported by Kimura *et al.* in the early 2000s [27, 28]. Despite these investigations, no existing physical model fully explains these behaviors.

In this work, we investigate the mechanisms behind the $T_C$ enhancement in Li-substituted $KNbO_3$, utilizing first-principles density functional theory (DFT) calculations. Our study focuses on the energy variations caused by the interplay between A-site and B-site displacements and changes in the chemical environment due to Li substitution at the A-site. We then summarize the impact of Li substitution on $T_C$ from multiple perspectives, aiming to clarify the origins of the observed increase. Experimental characterizations of materials are also provided to validate the simulation results.

## 2. Result and Discussion

2.1 Increased Curie temperature in Li-substituted KNbO$_3$

The preparation of KNbO$_3$ ceramics using conventional sintering methods is challenging due to limited densification and high sensitivity to moisture [29]. High porosity in sintered ceramics can often result in measurement artifacts and even electrical breakdown under low electric fields [30]. In this study, high-quality KNbO$_3$ and Li-substituted KNbO$_3$ (5 mol% of Li) ceramics were prepared using hot pressing. Figure 1 presents a comprehensive experimental comparison of these hot-pressed samples.

As revealed by the X-ray diffraction (XRD) pattern in Figure 1(a), the KNbO$_3$ sample exhibits a pure perovskite structure with an orthorhombic symmetry (*Amm2*) [31] and no impurity phase is detected. The Li-substituted KNbO$_3$ sample also possesses a similar orthorhombic crystal structure but is accompanied by a slight non-perovskite secondary phase. A quick peak fitting analysis performed on the perovskite 200$_{pc}$ peak around 45° suggests that both samples have very similar lattice parameters, though Li-substituted KNbO$_3$ seems slightly smaller (see Figure S1 and Tables S1-S2). A similar trend can be found in Li-substituted (K$_{0.5}$Na$_{0.5}$)NbO$_3$ ceramic [32].

The hot-pressed Li-substituted KNbO$_3$ ceramics exhibit significantly low porosity compared to ceramics prepared via conventional sintering, as demonstrated in the scanning electron microscopy (SEM) image in Figure 1(b). The back-scattered electron (BSE) signal shows contrast among certain grains, which likely correspond to the secondary phase evidenced in the XRD pattern. Energy-dispersive X-ray spectroscopy (EDS) analysis (see Figure S2) indicates that the brighter regions in the BSE image are potassium-deficient. XRD and SEM analyses suggest that the solid solubility of Li in the $KNbO_3$ matrix is limited, leading to the formation of the secondary phase. For reference, the solid solubility of Li in $(K_{0.5}Na_{0.5})NbO_3$ ceramic is close to 8 mol% [32], which seems slightly higher than $KNbO_3$.

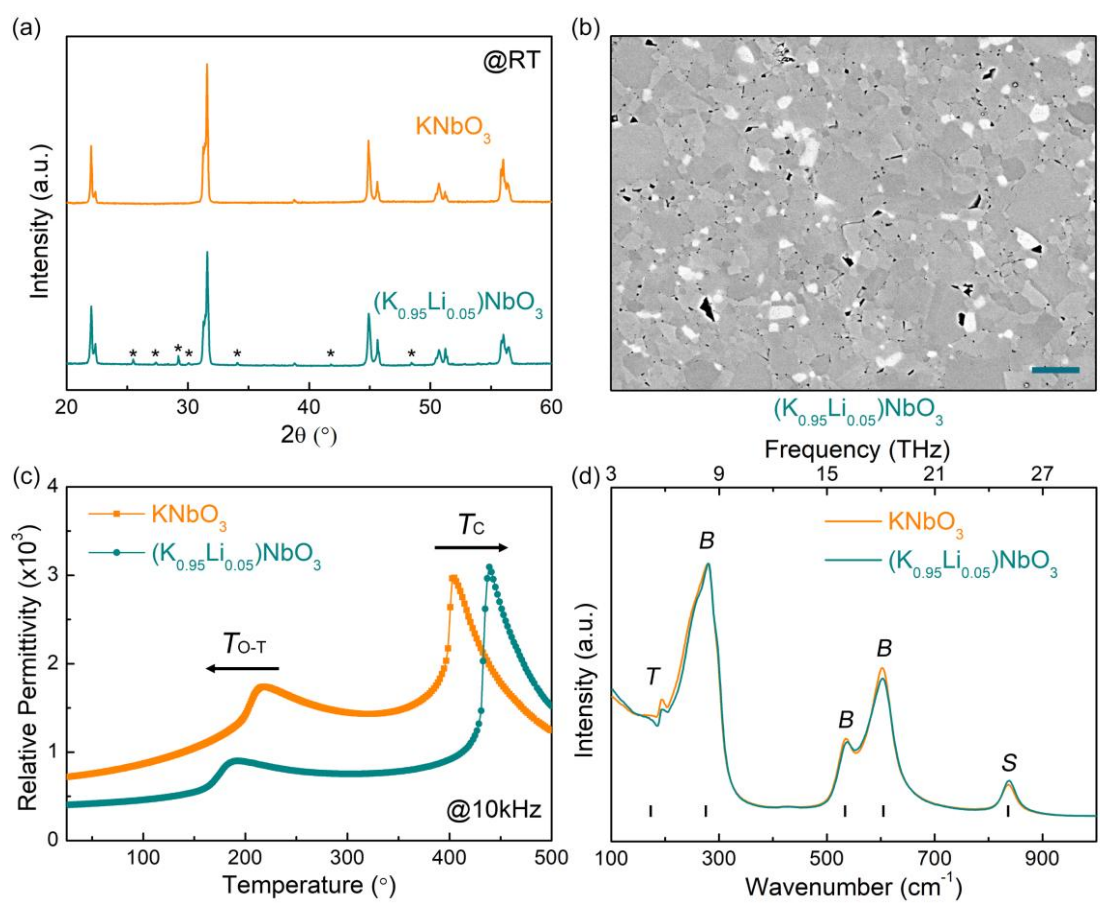

Figure 1. Experimental characterizations of hot-pressed $KNbO_3$ and Li-substituted $KNbO_3$ ceramics. (a) XRD measurement at room temperature. The signal from the secondary phase is marked with asterisks (*). (b) SEM back-scattered electron (BSE) image of as-sintered Li-substituted $KNbO_3$ sample. The scale bar is 5 μm. (c) Temperature-dependent relative permittivity measured at 10 kHz. (d) Raman spectra measured at room temperature. Symbols *T*, *B*, and *S* represent translational, bending, and stretching modes.

Figure 1(c) shows the temperature-dependent relative permittivity of both samples, revealing two significant peaks at ~402 °C and ~215 °C. The first peak corresponds to the ferroelectric tetragonal-to-paraelectric cubic phase transition ($T_C$) while the second peak corresponds to the ferroelectric orthorhombic-to-ferroelectric tetragonal phase transition ($T_{O-T}$). Li substitution increases $T_C$ to ~440 °C, while $T_{O-T}$ decreases to 193 °C.

Figure 1(d) and Figure S3 present the Raman spectra of both samples, showing significant overlap, which suggests a similar crystal structure and vibrational mode behavior. Based on the phonon spectrum of cubic $KNbO_3$ (Figure 2(a)), five Raman-active modes can be approximately identified at various wavenumbers. These include one stretching mode and three bending modes of the $NbO_6$ octahedron, as well as one translational mode involving the relative motion of $NbO_6$ against A-site atoms (K and Li) arranged from higher to lower wavenumbers. The peak coexistence in certain modes can be attributed to orthorhombic distortion in the crystal structure. Li substitution does not induce significant peak shifts of bending or stretching modes, indicating that the eigenvectors (representing atomic displacements) in these modes are largely unaffected. However, the spectral shape near the translational mode shows notable distortion, suggesting that $NbO_6$ exhibits different translational motions against A-site atoms between the pure $KNbO_3$ and Li-substituted $KNbO_3$. Unfortunately, the strong background signal at low wavenumbers hinders further analysis. Additional characterization details for both samples are provided in the supplemental file (Figures S4-S5).

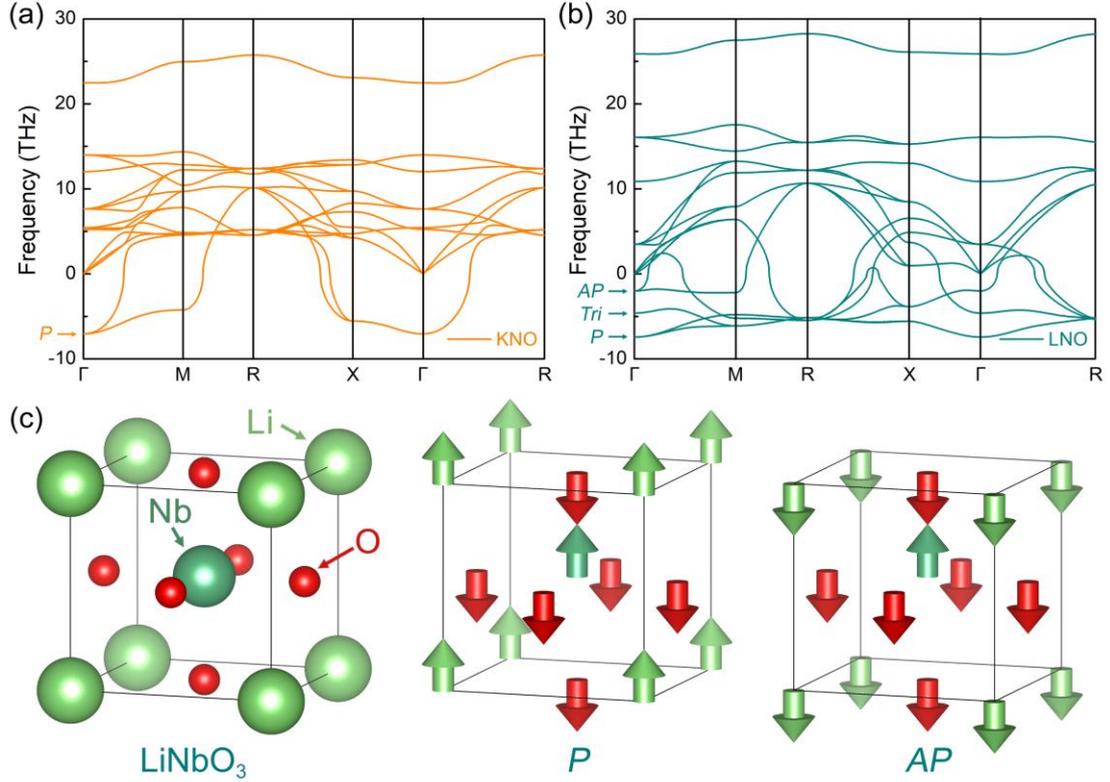

Figure 2. Phonon spectra of *pm-3m* cubic phase (a) KNbO$_3$ and (b) LiNbO$_3$. Symbols *P*, *AP*, and *Tri* correspond to the polar mode, anti-phase polar mode, and trigonal mode, respectively. (c) Schematics of the displacement directions of ions in polar (*P*) and anti-phase polar (*AP*) modes in LiNbO$_3$. Please note that the vectors do not reflect real displacement magnitude.

2.2 Ferroelectric instability and soft polar mode

From the phonon spectra of KNbO$_3$ (denoted as KNO) shown in Figure 2(a), it is shown that there is an imaginary mode with a frequency of -7.05 THz at Γ point, which corresponds to the soft polar mode that causes the ferroelectric instability. In this polar mode, which is an optical mode, the K and Nb cations move in the direction opposite to the O anion. However, there are three imaginary modes at Γ point for LiNbO$_3$ (denoted as LNO), *i.e.*, -2.01, -4.62, -7.43 THz. Akin to KNbO$_3$, the imaginary mode at -7.43 THz corresponds to the typical polar mode. Interestingly, the imaginary mode at -2.01 THz also corresponds to a polar mode, but the displacement vector of the Li cation becomes anti-phase to that of the Nb cation, herein denoted as anti-phase polar mode. The difference between the two modes is illustrated in Figure 2(c). Lastly, the imaginary mode at -4.62 THz should correspond to the structural instability leading to the experimentally observed ferroelectric phase in trigonal crystal structure [33]. The detailed eigendisplacements in each mode are in the supplemental file (Table S3-S6).

2.3 B-site driven ferroelectric instability

By constructing a 2x2x2 cubic *pm-3m* supercell, a certain number of K cations is homogenously replaced with Li cations to achieve different Li concentrations, including 0%, 12.5 %, 50%, 87.5%, and 100%, named KNO, KLN125, KLN500, KLN875, and LNO, respectively (see Figure S6 for atomic configurations). Experimentally, the solid solution limit for Li in niobate perovskites is usually around 5-10% [32], due to the difference in chemical nature between the Li cation and original A-site cations. Although the Li concentrations modeled in our simulations exceed these experimental limits, the qualitative trends revealed by the simulations remain relevant for understanding experimental observations.

To elucidate the ferroelectric behavior, it is essential to separately analyze the off-center displacements of the B-site and A-site cations, as they can contribute to ferroelectric instability in distinct ways. Figure 3 illustrates the energy variation associated with off-center displacements of B-site and A-site cations for structures with varying Li concentrations. In Figure 3(a), the energy variation of the polar mode by exclusively considering the off-center displacement of B-site cations is calculated. The resulting double-well energy profile indicates ferroelectric instability. As the Li concentration increases, the depth of the energy wells decreases, suggesting a reduction in ferroelectric instability.

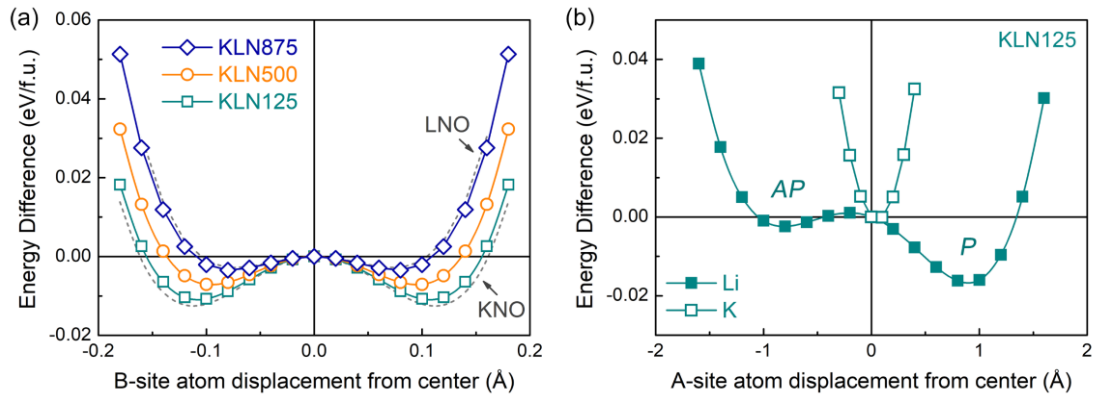

Figure 3. (a) Energy variation as a function of the off-center displacement of the B-site cation in different compositions. (b) The energy variation as a function of the off-center displacement of the A-site cation with pre-fixed B-site cation displaced to ~0.1Å in KLN125. *P* and *AP* correspond to polar and anti-phase polar modes, respectively.

According to Landau's phenomenological theory, the simulated smaller energy difference between the saddle point (paraelectric phase) and the minimum point (ferroelectric phase) should lead to a lower $T_C$, which intuitively appears contradictory to the experimental result. This apparent discrepancy highlights the need for further investigation into the compensating mechanisms that contribute to the experimentally observed increase in $T_C$ with Li substitution.

2.4 Compressive chemical pressure induced by Li

Ferroelectric instability is typically related to hybridization among cations and anions [31], which can be analyzed through the Born effective charges (BEC) [34]. The BEC of all elements in different compositions are summarized in Table 1. In the KNO structure, the BECs of Nb and O ions deviate strongly from their nominal values of +5 and -2, respectively, highlighting strong hybridization. In contrast, the BECs of K and Li remain close to their nominal values of +1, indicating predominantly ionic behavior. Further details on this topic can be found in our previous work [31]. As the Li concentration increases, the slightly reduced BECs of Nb and O ions, reflecting a weaker Nb–O hybridization, can reasonably explain the observation of diminished ferroelectric instability in Figure 2(a).

Table 1. Born effective charges (unit in $e$) in different compositions. Please note that the charges are averaged over all present ions in the supercell of the same element.

| Compositions | K | Li | Nb | $O_\parallel$ | $O_\perp$ |
|---|---|---|---|---|---|
| KNO | 1.14 | N/A | 9.75 | -7.42 | -1.74 |
| KLN125 | 1.14 | 1.25 | 9.66 | -7.36 | -1.69 |
| KLN500 | 1.12 | 1.26 | 9.57 | -7.34 | -1.68 |
| KLN875 | 1.14 | 1.25 | 9.48 | -7.31 | -1.67 |
| LNO | N/A | 1.25 | 9.46 | -7.31 | -1.70 |

Interestingly. the BECs of the K and Li cations remain nearly unchanged, suggesting that these cations do not directly influence the chemical bonding in these compositions. However, due to the smaller ionic radius of Li (1.13±0.17 Å by extrapolation on existing data, see Table S7) compared to K (1.64 Å) [35], Li substitution is expected to shrink the unit cell, generating compressive stress often referred to as chemical pressure [36]. The calculated unit cell volume as a function of Li concentration, summarized in Table 2, aligns well with this expectation and corroborates chemical intuition.

Table 2. Lattice parameters and volume of normalized cubic unit cell of different compositions.

| Parameters | KNO | KLN125 | KLN500 | KLN875 | LNO |
|---|---|---|---|---|---|
| $a=b=c$ (Å) | 4.029 | 4.016 | 3.991 | 3.960 | 3.953 |
| Volume (Å$^3$) | 65.41 | 64.77 | 63.57 | 62.10 | 61.77 |

Inspired by this volume variation, we further examine the influence of the compressive chemical pressure induced by Li on BECs. Hydrostatic compression was applied to the unsubstituted KNO structure to mimic the chemical pressure, and the resulting BECs are presented in Table 3. Under increasing compressive stress, the BECs of Nb and O decrease slightly, whereas those of K remain nearly unchanged. This indicates that compressive stress can indeed influence the Nb–O hybridization, a finding consistent with experimental observations that applied compressive stress lowers $T_C$ [37]. Inferring from this, we believe that reduced ferroelectric instability driven by B-site atoms originates from the compressive chemical pressure induced by Li.

Table 3. Born effective charges (unit in $e$) in KNO under different compressive strains.

| Volume (Å$^3$) | Pressure (GPa) | K | Nb | $O_\parallel$ | $O_\perp$ |
|---|---|---|---|---|---|
| 65.41 | 0.05 | 1.14 | 9.75 | -7.42 | -1.74 |
| 64.92 | 1.33 | 1.14 | 9.73 | -7.40 | -1.74 |
| 64.43 | 2.72 | 1.14 | 9.72 | -7.38 | -1.74 |
| 63.95 | 4.15 | 1.14 | 9.70 | -7.36 | -1.74 |
| 63.47 | 5.64 | 1.14 | 9.69 | -7.34 | -1.74 |
| 63.00 | 7.18 | 1.14 | 9.67 | -7.32 | -1.74 |

2.4 A-site driven ferroelectric instability

Phonon spectra comparisons between KNO and LNO reveal that the Li cation exhibits a substantially larger off-center displacement than the K cation (Figure 2). To further explore this behavior, we simulated the off-center displacements of the A-site cations in the KLN125 structure while pre-fixing the Nb cations at an off-center displacement of ~0.1 Å with minimized energy. The results, shown in Figure 3(b), provide several notable insights.

First, the energy minimum for the K cation is near zero, indicating a limited tendency for off-center displacement. However, it shows a slightly positive value due to coupling with the off-center Nb cation. Second, the Li cation exhibits two distinct energy minima corresponding to enormous off-center displacements compared to K. On the positive side, a deep energy well of approximately 0.016 eV/f.u. is observed, while on the negative side, a shallower energy well of approximately 0.002 eV/f.u. is present. The coexistence of these minima can be related to two distinct polar modes in the phonon spectrum of LNO, occurring at -2.01 THz and -7.43 THz, respectively.

2.5 Impact on $T_C$

Clarity emerges when the interplay between the A-site and B-site contributions is considered. This substantial off-center displacement of the Li cation not only compensates for the reduced energy depth observed in the B-site-cation-determined energy curve but also deepens the energy well further. In Figure 4(a), the energy differences between the paraelectric and ferroelectric phases across various compositions are calculated, revealing a nearly linear amplification with increased Li concentration. Figure 4(b) illustrates a schematic representation of the energy curve's evolution with Li substitution. The pronounced off-center displacement of the Li cation deepens the overall energy well, thus requiring greater thermal energy for the system to transition from the ferroelectric phase to the paraelectric phase. Namely, the enhanced soft polar mode, which deepens the energy well, contributes directly to an increase in $T_C$.

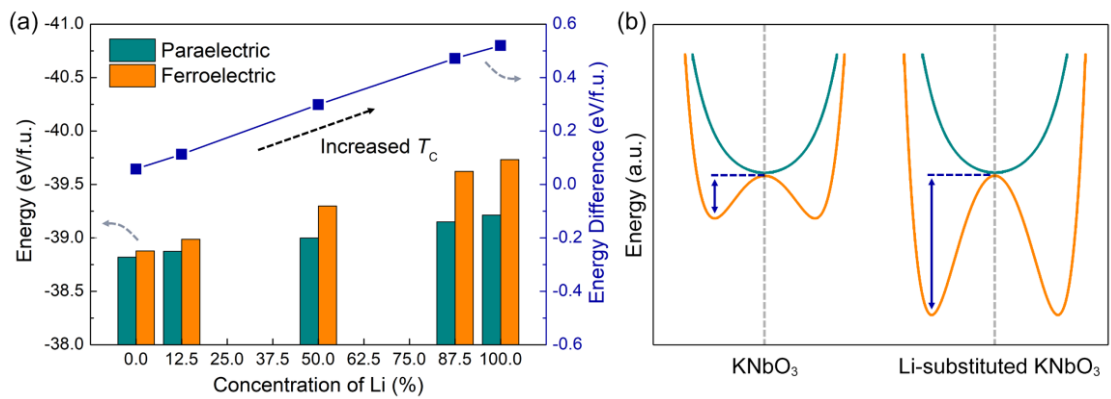

Figure 4. (a) Energy difference between paraelectric and ferroelectric phases in different compositions. (b) Illustration of the energy curve variation in the unsubstituted and Li-substituted KNbO$_3$. The illustration might be exaggerated to show the difference.

Notably, trigonal LNO exhibits an exceptionally high $T_C$ ~1500K [38]. The energy difference between the paraelectric and ferroelectric phases in LNO is calculated to be 520 meV/f.u., nearly nine times higher than that in KNO (58 meV/f.u.), which corresponds to an experimentally observed $T_C$ ~700K [39]. These findings demonstrate that examining the energy curve variations induced by chemical substitution provides a robust qualitative framework for predicting changes in $T_C$. The near-linear relationship between energy difference and Li concentration also facilitates $T_C$ prediction for interpolated compositions, especially those within the solid solubility limit.

2.6 Impact on $T_{O\text{-}T}$

Intriguingly, while the $T_C$ increases with Li substitution, $T_{O\text{-}T}$ decreases. Typically, phase transition temperatures follow a similar trend, as observed under externally applied pressure [37]. To understand this peculiarity, we analyzed the enhanced ferroelectric instability caused by the off-center displacement of the Li cation along polarization vectors corresponding to different phases: <001> for tetragonal, <011> for orthorhombic, and <111> for rhombohedral.

Figure 5 shows the energy variation as a function of the off-center displacement of the Li cation in cubic KLN125 along these directions. Notably, Li can displace significantly further along <001> (~0.9 Å) than along <011> and <111> (~0.6 Å), where the larger displacement also contributes to a deeper energy well. The result suggests Li substitution preferentially stabilizes the tetragonal phase over the orthorhombic and rhombohedral phases. Figures 5(b1)-(b4) depict the environment of an A-site atom in a perovskite structure. While the B-site atom usually coordinates with six O anions, the A-site atom coordinates with 12 O anions. Thus, A-site usually has a larger space intrinsically. The smaller ionic radius of Li compared to K allows it to move further off-center, but its displacement is constrained along certain directions. For instance, Li can collide with the O anions at shorter distances along <011> and <111> directions. This behavior contrasts with pure $KNbO_3$, where K remains near the center, contributing minimally to phase stability. The result suggests that the stabilization of the tetragonal phase over the orthorhombic phase is the main reason for the decreased $T_{O\text{-}T}$ upon Li substitution.

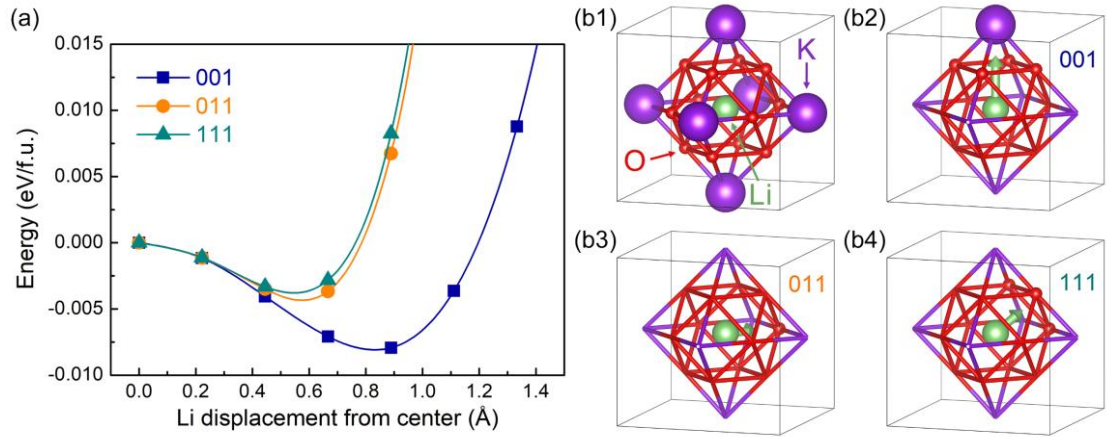

Figure 5. (a) Energy variation as a function of the off-center displacement of the Li cation in cubic KLN125 along different directions. (b1) A schematic of the Li cation surrounded by the nearest O anions and K cations. Schematics of the Li cation displaced along (b2) <001>, (b3) <011>, and (b4) <111> directions. The displacement vector is referenced based on the displacements with the lowest energy shown in (a) and amplified 3 times for illustration. Some nearest atoms along the displacement are also displayed in these figures to guide the eye.

A parallel can be drawn with $PbTiO_3$, where the ferroelectric tetragonal phase is also energetically favorable. In this case, Pb exhibits a large off-center displacement in the tetragonal phase [40], due to the strong hybridization between Pb and O [41]. In contrast, Ba in $BaTiO_3$ exhibits a small off-center displacement, akin to K in $KNbO_3$ [40], allowing orthorhombic and rhombohedral phases to exist at lower temperatures [42]. We also noted that the interplay between A-site and B-site driven ferroelectricity was also investigated in the $(Ba, Ca)(Ti, Zr)O_3$ ferroelectric solid solution [43, 44]. Such evidence suggests that the displacement of the A-site cation is as crucial as that of the B-site cation in determining phase stability in ferroelectric perovskites.

2.7 Metastable anti-phase polar state

The anti-phase displacement between the Li and Nb cations is evident in the LNO phonon spectrum and also reflected in the double minima of the Li-displacement energy curve for KLN125 (Figure 3b). The anti-phase polar mode in KLN125 was further relaxed under tetragonal symmetry, resulting in a higher-energy anti-phase polar structure (~52 meV/f.u.) compared to the polar structure, as shown in Figure 6(a). It is hypothesized that the anti-phase polar structure may serve as an intermediate state between the polar and paraelectric structures.

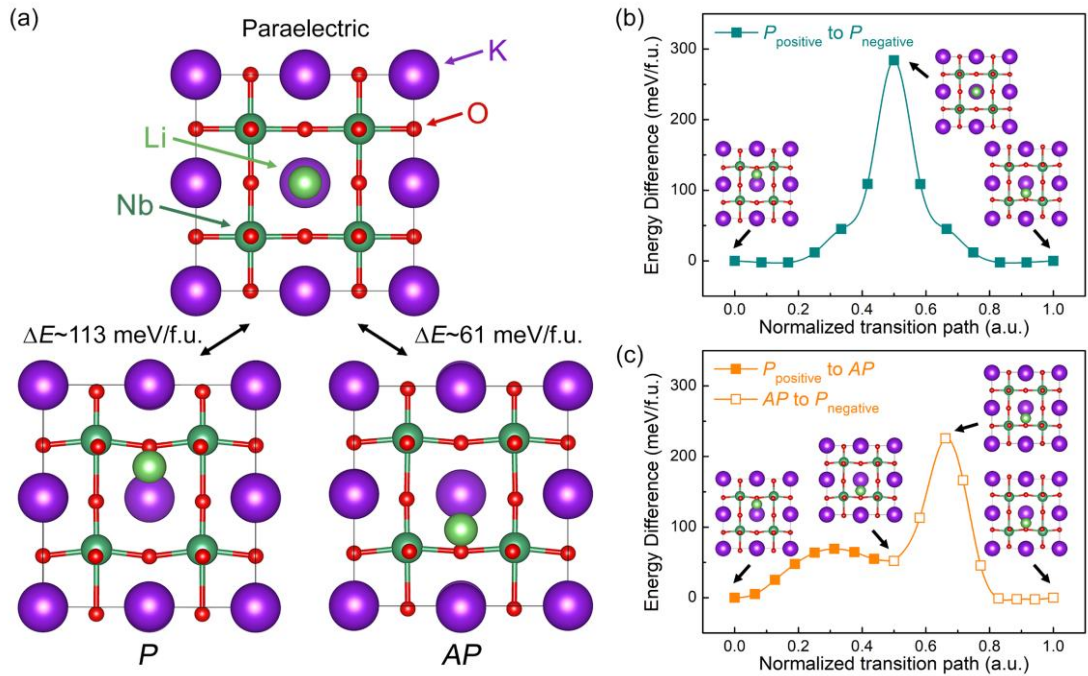

Figure 6. (a) Relaxed configurations of paraelectric phase, ferroelectric phase with polar and anti-phase polar modes in KLN125. Transition path of the polarization switching between the ferroelectric phase with positive polarization and the ferroelectric phase with negative polarization (b) via paraelectric phase and (c) via anti-phase polar phase in KLN125.

To test this hypothesis, the polarization switching transition path via the paraelectric structure was calculated using the nudged elastic band method (Figure 6b). The results show that an energy barrier of ~284 meV/f.u. is required for polarization switching through the ideal paraelectric structure. Interestingly, two additional shoulders near the energy barrier suggest the existence of a previously unconsidered intermediate structure along the transition path. To explore this, the polarization switching path via the anti-phase polar structure was examined (Figure 6c).

In the normalized transition path (0 to 0.5), a small energy barrier corresponds to the Li cation switching from positive to negative off-center displacement, while the other ions remain nearly fixed. From 0.5 to 1, the polarization switching is primarily driven by off-center displacements of the Nb and O ions. The maximum energy barrier for switching via the anti-phase polar structure is ~226 meV/f.u., notably lower than the 284 meV/f.u. required for the transition through the paraelectric structure, indicating the anti-phase polar structure is energetically favorable as an intermediate state. The results also suggest that the anti-phase polar structure could exist near 180° domain walls where the polarization reversal occurs.

Although the anti-phase polar structure is metastable, it is suggested that it could be observed experimentally under certain conditions, *e.g.*, located at domain walls stabilized by neighboring domains, applied electric field, or other specific thermodynamic conditions. However, its stability is highly sensitive to ambient thermal energy, as the reverse transition energy barrier from the anti-phase polar structure to the in-phase polar structure is only ~17 meV/f.u., rendering it volatile under standard conditions. Stabilizing the anti-phase polar structure at room temperature could theoretically result in a lower $T_C$.

## 3  Conclusions

In summary, the impact of Li substitution on the enhanced $T_C$ in KNbO$_3$ was investigated using DFT calculations and validated with experimental characterization. While the ferroelectric instability contributed by Nb–O hybridization diminishes with increasing Li concentration due to compressive chemical pressure, the pronounced off-center displacement of the Li cation compensates for this reduction. This displacement deepens the energy well, thereby reinforcing ferroelectric instability and leading to a predicted increase in $T_C$ in Li-substituted KNbO$_3$. Additionally, the large displacement of the Li cation preferentially stabilizes the tetragonal phase over the orthorhombic phase, resulting in a decrease in $T_{O\text{-}T}$. Another intriguing finding is the identification of a metastable anti-phase polar state in Li-substituted KNbO$_3$. During polarization switching, the independent motion of the Li and Nb cations is energetically more favorable than their simultaneous displacement. Under specific conditions, this metastable anti-phase polar state could manifest, potentially changing $T_C$. These results provide valuable insights into the fundamental mechanisms behind the increased $T_C$ in ferroelectric niobate-based perovskites, offering a clear framework for materials engineering and the design of high-performance ferroelectric materials for high-temperature applications.

# 4 Experimental and Calculation details

4.1 Preparation and characterization of ceramic samples

The pure $KNbO_3$ and Li-substituted $KNbO_3$ ($K_{0.95}Li_{0.05}NbO_3$) ceramic samples are prepared using hot pressing to ensure high density. $K_2CO_3$ (>99%, Sinopharm), $Li_2CO_3$ (>99%, Sinopharm), and $Nb_2O_5$ (>99.99%, Sinopharm) are used as the raw powders and ball-milled for 24 h before calcination at 930 °C for 4 h via the conventional solid-state reaction route. The calcined powder is hot-pressed for 2 h under a uniaxial pressure of 30 MPa at 920 °C. The hot-pressed sample is cut and polished, before annealing at 880 °C in an air atmosphere for 12 h.

The crystal structures of the final products are characterized using XRD with Cu Kα radiation (D/max2500, Rigaku) with a 2θ step of 0.02°. The Raman spectra are measured using Raman spectroscopy (HR800, HORIBA) with an excitation wavelength of 633 nm at 5 random spots in each ceramic sample to ensure consistency. The final spectra are averaged from all measurements and normalized according to the highest peak for qualitative comparison, where the original data can be found in the supplemental file. SEM (JSM-6460LV, JEOL) is used to characterize the morphology of sintered ceramics. The temperature-dependent dielectric properties are measured using an impedance analyzer (TH2827, Changzhou Tonghui Electronic Co). For this measurement, the samples are coated with silver paste and annealed at 600°C for 30 min to form electrodes.

## 4.2 First-principles calculations

First-principles density functional theory (DFT) calculation is performed using VASP software. Generalized gradient approximation (GGA) with the Perdew–Burke–Ernzerhofer (PBE) exchange-correlation function is used. The valence electrons considered in four types of elements are $3s^23p^64s^1$ (K), $1s^22s^1$ (Li), $4s^24p^65s^14d^4$ (Nb), and $2s^22p^4$ (O). Plane-wave cutoff energy is set to 500 eV. The convergence criteria for energy and force are set to $10^{-6}$ eV and 0.005 eV/Å, respectively. Γ-center K-point meshes with a spacing of 0.12 Å$^{-1}$ are created for all calculated structures.

The phonon spectra are calculated using the finite displacement method, and the non-analytical term correction is implemented using PHONOPY [45]. A 3x3x3 supercell is created for phonon calculation. The software automatically estimates band connection with 101 sampling points along each path. The born effective charges are calculated using density functional perturbation theory (DFPT). The transition paths are computed using the climbing nudged elastic band (c-NEB) method based on the VASP transition state theory (VTST) tool [46]. Visualization of atomic configurations in figures is achieved by using VESTA [47].

**Acknowledgment**

This work was supported by the National Natural Science Foundation of China (No. W2433118) and the Beijing Natural Science Foundation (No. IS24026). Fang-Zhou Yao appreciates the support from the National Natural Science Foundation of China (No. U22A20254). Yan Wei appreciates the support from the National Natural Science Foundation of China (No. 82225012, 82430033). Shi-Dong Wang appreciates the support from Beijing Natural Science Foundation (No. JQ24049, L222066). We sincerely appreciate the insightful discussion with our colleague, Dr. Yi-Xuan Liu.

# References

1. Zhou X, Zhang J, Luo H, et al. Nature Communications. 2024;15(1):6625.

2. Guo J, Wang Z, Tian G, et al. Journal of Materiomics. 2025;11(1):100830.

3. Wang H, Guan Z, Li J, et al. Advanced Materials. 2024;36(15):2211305.

4. Qi H, Chen L, Deng S, et al. Nature Reviews Materials. 2023;8(6):355-6.

5. Jiang Y, Zhang M-H, Wu C-F, et al. Nature Communications. 2024;15(1):9024.

6. Qiu C, Wang B, Zhang N, et al. Nature. 2020;577(7790):350-4.

7. Go S-H, Park S-J, Kim S-H, et al. Journal of Materiomics. 2024;10(3):632-42.

8. Hellenbrand M, Teck I, MacManus-Driscoll JL. MRS Communications. 2024:1-14.

9. Wang Y, Tao L, Guzman R, et al. Science. 2023;381(6657):558-63.

10. Liu Z, Wu H, Yuan Y, et al. Current Opinion in Solid State and Materials Science. 2022;26(5):101016.

11. Dong Y, Zou K, Liang R, et al. Progress in Materials Science. 2023;132:101026.

12. Liu L, Yi J, Tang M, et al. Advanced Functional Materials. 2024;34(21):2314807.

13. Hua Y, Qian J, Yang Y, et al. Advanced Functional Materials. 2024:2414348.

14. Wang B, Huangfu G, Zheng Z, et al. Advanced Functional Materials. 2023;33(20):2214643.

15. Zheng T, Yu Y, Lei H, et al. Advanced Materials. 2022;34(8):2109175.

16. Cen Z, Feng M, Cao F, et al. Journal of Materiomics. 2024;10(6):1165-75.

17. Dove MT. Introduction to lattice dynamics: Cambridge University Press; 1993.

18. Yu Z, Fan N, Fu Z, et al. Nature Communications. 2024;15(1):3438.

19. Zhang H, Thong H-C, Bastogne L, et al. Physical Review B. 2024;110(5):054109.

20. Ghosez P, Cockayne E, Waghmare U, et al. Physical Review B. 1999;60(2):836.

21. Zhang H, Chao C-H, Bastogne L, et al. Physical Review B. 2023;108(14):L140304.

22. Ghosez P, Gonze X, Michenaud J-P. Europhysics Letters. 1996;33(9):713.

23. Li JF, Wang K, Zhu FY, et al. Journal of the American Ceramic Society. 2013;96(12):3677-96.

24. Saito Y, Takao H, Tani T, et al. Nature. 2004;432(7013):84-7.

25. Machado R, Sepliarsky M, Stachiotti M. Physical Review B. 2012;86(9):094118.

26. Li CBW, Thong HC, Liu YX, et al. Advanced Functional Materials. 2022;32(38):2204421.

27. Kimura M, Ogawa T, Ando A, et al. Piezoelectric properties of metastable (Li, Na)NbO3 ceramics. Proceedings of the 13th IEEE International Symposium on Applications of Ferroelectrics, 2002 ISAF 2002; 2002: IEEE.

28. Kimura M, Ando A, Shiratsuyu K, et al. Transactions of the Materials Research Society of Japan. 2004;29(4):1049.

29. Kakimoto K-i, Masuda I, Ohsato H. Journal of the European Ceramic Society. 2005;25(12):2719-22.

30. Birol H, Damjanovic D, Setter N. Journal of the American Ceramic Society. 2005;88(7):1754-9.

31. Thong H-C, Xu B, Wang K. Applied Physics Letters. 2022;120(5).

32. Guo Y, Kakimoto K-i, Ohsato H. Applied Physics Letters. 2004;85(18):4121-3.

33. Gopalan V, Dierolf V, Scrymgeour DA. Annual Review of Materials Research.

2007;37(1):449-89.

34. Ghosez P, Michenaud J-P, Gonze X. Physical Review B. 1998;58(10):6224.

35. Shannon RD. Foundations of Crystallography. 1976;32(5):751-67.

36. Lin K, Li Q, Yu R, et al. Chemical Society Reviews. 2022;51(13):5351-64.

37. Zhou Z, Fang L, Xiong Z, et al. Applied Physics Letters. 2023;123(1).

38. Gopalan V, Sanford NA, Aust J, et al. Handbook of Advanced Electronic and Photonic Materials and Devices: Elsevier; 2001. p. 57-114.

39. Fontana M, Metrat G, Servoin J, et al. Journal of Physics C: Solid State Physics. 1984;17(3):483.

40. Warren W, Robertson J, Dimos D, et al. Physical Review B. 1996;53(6):3080.

41. Kuroiwa Y, Aoyagi S, Sawada A, et al. Physical Review Letters. 2001;87(21):217601.

42. Zhang J, Bastogne L, He X, et al. Physical Review B. 2023;108(13):134117.

43. Amoroso D, Cano A, Ghosez P. Physical Review B. 2018;97(17):174108.

44. Amoroso D, Cano A, Ghosez P. Applied Physics Letters. 2019;114(9).

45. Togo A, Chaput L, Tadano T, et al. Journal of Physics: Condensed Matter. 2023;35(35):353001.

46. Sheppard D, Xiao P, Chemelewski W, et al. The Journal of Chemical Physics. 2012;136(7):074103.

47. Momma K, Izumi F. Journal of Applied Crystallography. 2011;44(6):1272-6.

Supplementary Materials for

# Origin of Increased Curie Temperature in Lithium-Substituted Ferroelectric Niobate Perovskite: Enhancement of the Soft Polar Mode

Hao-Cheng Thong,[1] Fang-Zhou Yao,[2,*] Xian-Xian Cai,[1] Ze Xu,[1] Mao-Hua Zhang,[2] Huazhang Zhang,[3] Ben Xu,[4] Yan Wei,[5,*] Shi-Dong Wang,[6,*] and Ke Wang[1]

[1] State Key Laboratory of New Ceramics and Fine Processing, School of Materials Science and Engineering, Tsinghua University, Beijing 100084, P. R. China

[2] Research Center for Advanced Functional Ceramics, Wuzhen Laboratory, Jiaxing 314500, P. R. China

[3] School of Physics and Mechanics, Wuhan University of Technology, Wuhan 430070, P. R. China

[4] Graduate School, China Academy of Engineering Physics, Beijing 100193, P. R. China

[5] Department of Geriatric Dentistry; NMPA Key Laboratory for Dental Materials National Engineering Laboratory for Digital and Material Technology of Stomatology, Peking University School and Hospital of Stomatology, Beijing 100081, P. R. China

[6] Musculoskeletal Tumor Center, Peking University People's Hospital, Beijing 100044, P. R. China

* Corresponding author: Fang-Zhou Yao (yaofangzhou@xjtu.edu.cn); Yan Wei (kqweiyan@bjmu.edu.cn); Shi-Dong Wang (bonetumor@bjmu.edu.cn)

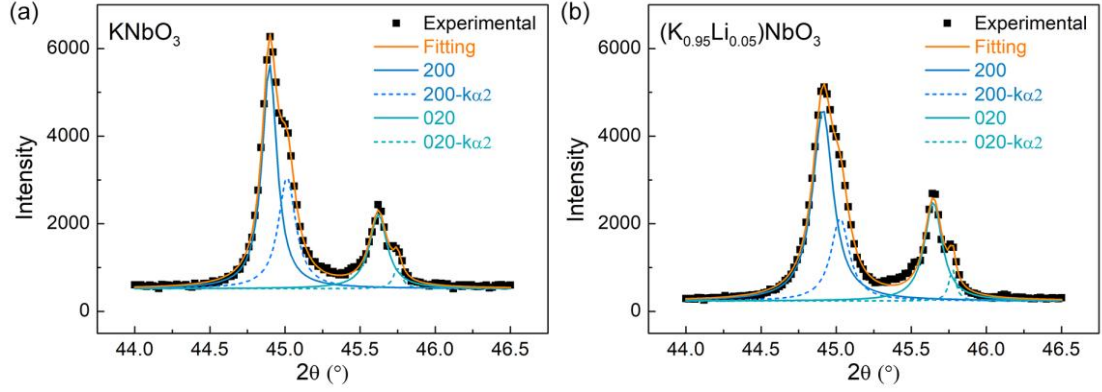

Figure S1. Peak fitting of 200pc peaks of orthorhombic *amm2* phase in (a) $KNbO_3$ and (b) $(K_{0.95}Li_{0.05})NbO_3$ samples. Peak fitting is performed using Lorentz distribution.

Table S1. Summary of the fitting result.

|  | Center of 200 (Error) | Center of 020 (Error) | Reduced $\chi^2$ |
|---|---|---|---|
| $KNbO_3$ | 44.8970 (0.0008) | 45.6212 (0.0019) | 0.9979 |
| $(K_{0.95}Li_{0.05})NbO_3$ | 44.9108 (0.0020) | 45.6461 (0.0017) | 0.9963 |

Table S2. Estimated lattice parameters from the fitting result in Table S1.

|  | *a=c* (Error) | *b* (Error) |
|---|---|---|
| $KNbO_3$ | 4.0345 (0.0001) | 3.9738 (0.0002) |
| $(K_{0.95}Li_{0.05})NbO_3$ | 4.0333 (0.0002) | 3.9718 (0.0001) |

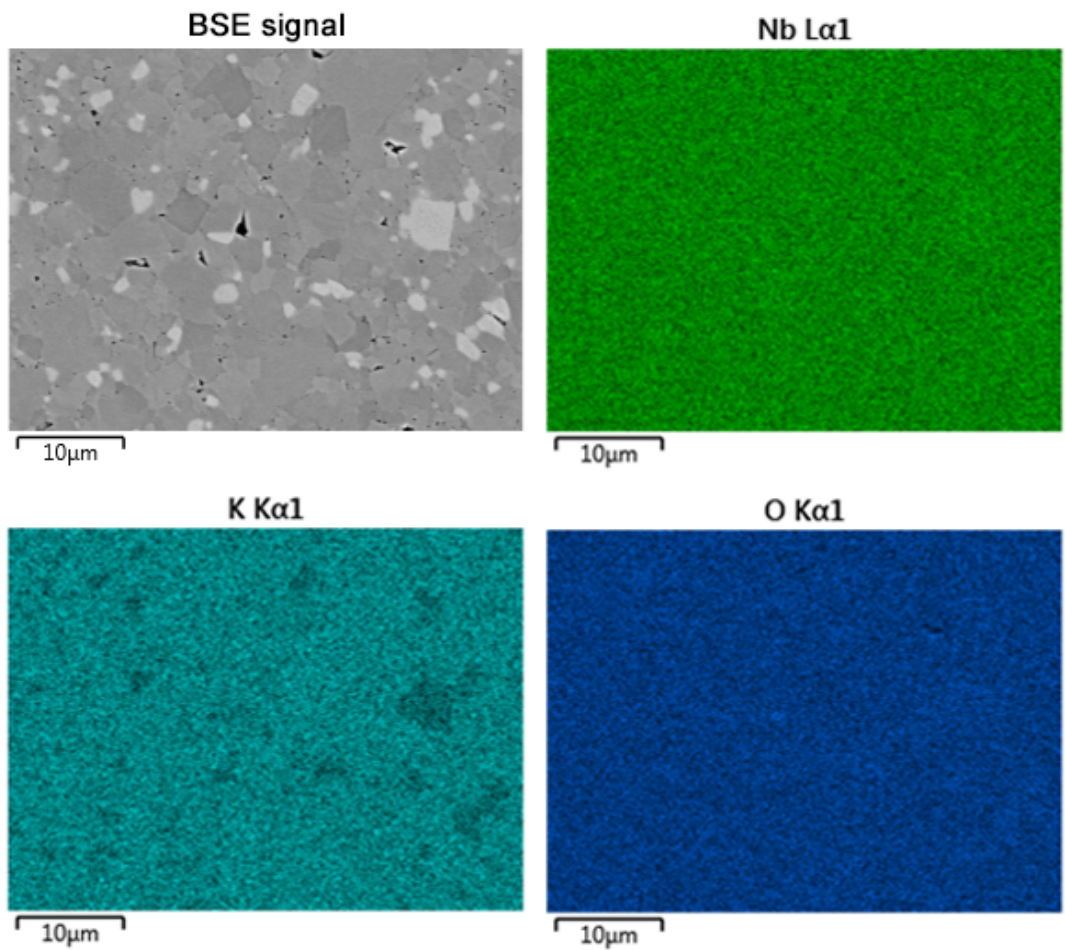

Figure S2. EDS analysis of as-sintered $(K_{0.95}Li_{0.05})NbO_3$ ceramic samples.

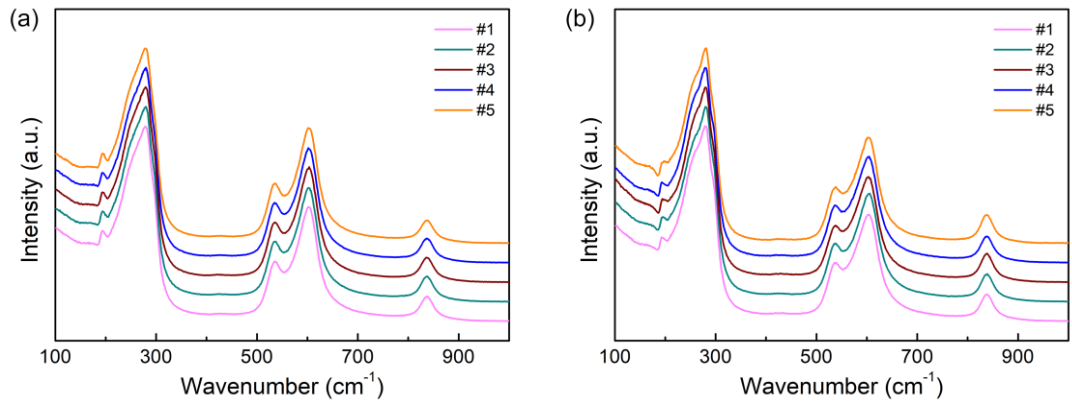

Figure S3 Raman spectra of (a) $KNbO_3$ and (b) $(K_{0.95}Li_{0.05})NbO_3$ ceramic samples.

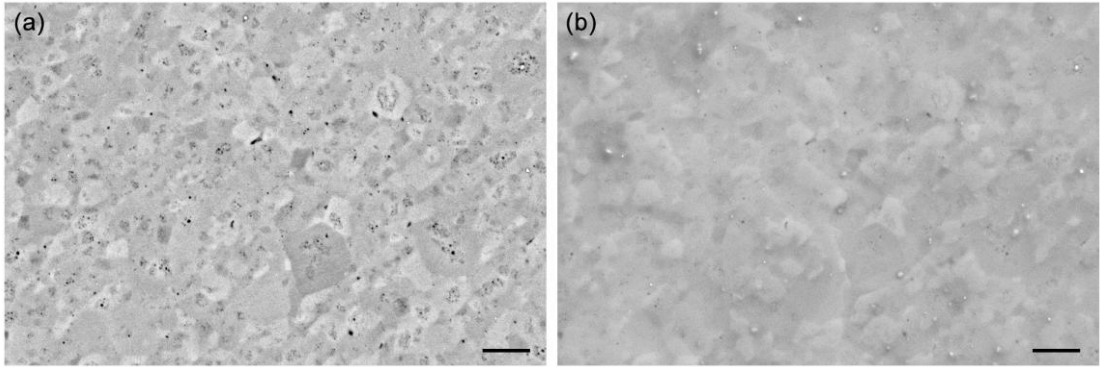

Figure S4. SEM (a) secondary-electron signal and (b) back-scattered electron signal of undoped KNbO$_3$ sample. The scale bar is 2μm.

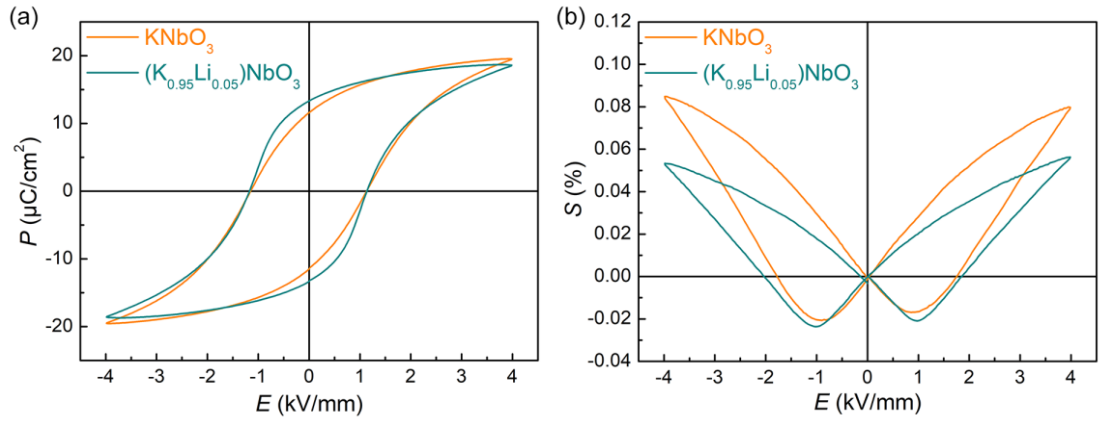

Figure S5. (a) Polarization-electric field and (b) strain-electric field measurements of $KNbO_3$ and $(K_{0.95}Li_{0.05})NbO_3$ ceramic samples.

Table S3. Eigendisplacements of atoms of the soft polar mode (frequency at -7.05 THz) in KNbO$_3$ at Γ point. The displacement vectors are re-aligned to [001].

|     | x (Å)  | y (Å)  | z (Å)  | Magnitude (Å) |
| --- | ------ | ------ | ------ | ------------- |
| Nb  | 0.000  | 0.000  | 0.061  | 0.061         |
| K   | 0.000  | 0.000  | -0.006 | 0.006         |
| O$_1$ | 0.004  | -0.004 | -0.151 | 0.151         |
| O$_2$ | -0.004 | 0.000  | -0.095 | 0.095         |
| O$_3$ | 0.000  | 0.004  | -0.095 | 0.095         |

Table S4. Eigendisplacements of atoms of the soft polar mode (frequency at -7.43 THz) in LiNbO$_3$ at Γ point. The displacement vectors are re-aligned to [001].

|     | x (Å) | y (Å) | z (Å)  | Magnitude (Å) |
| --- | ----- | ----- | ------ | ------------- |
| Nb  | 0.000 | 0.000 | 0.030  | 0.030         |
| Li  | 0.000 | 0.000 | 0.257  | 0.257         |
| O$_1$ | 0.000 | 0.000 | -0.074 | 0.074         |
| O$_2$ | 0.000 | 0.000 | -0.107 | 0.107         |
| O$_3$ | 0.000 | 0.000 | -0.107 | 0.107         |

Table S5. Eigendisplacements of atoms of the soft anti-phase polar mode (frequency at -2.01 THz) in LiNbO$_3$ at Γ point. The displacement vectors are re-aligned to [001].

|     | x (Å) | y (Å) | z (Å)  | Magnitude (Å) |
| --- | ----- | ----- | ------ | ------------- |
| Nb  | 0.000 | 0.000 | 0.055  | 0.055         |
| Li  | 0.000 | 0.000 | -0.265 | 0.265         |
| O$_1$ | 0.000 | 0.000 | -0.074 | 0.074         |
| O$_2$ | 0.000 | 0.000 | -0.068 | 0.068         |
| O$_3$ | 0.000 | 0.000 | -0.068 | 0.068         |

Table S6. Eigendisplacements of atoms of the soft mode (frequency at -4.62 THz) in LiNbO$_3$ at Γ point.

|     | x (Å)  | y (Å)  | z (Å) | Magnitude (Å) |
| --- | ------ | ------ | ----- | ------------- |
| Nb  | -0.017 | -0.017 | 0.000 | 0.024         |
| Li  | 0.258  | 0.258  | 0.000 | 0.364         |
| O$_1$ | -0.015 | -0.015 | 0.000 | 0.021         |
| O$_2$ | -0.015 | 0.017  | 0.000 | 0.023         |
| O$_3$ | 0.017  | -0.015 | 0.000 | 0.023         |

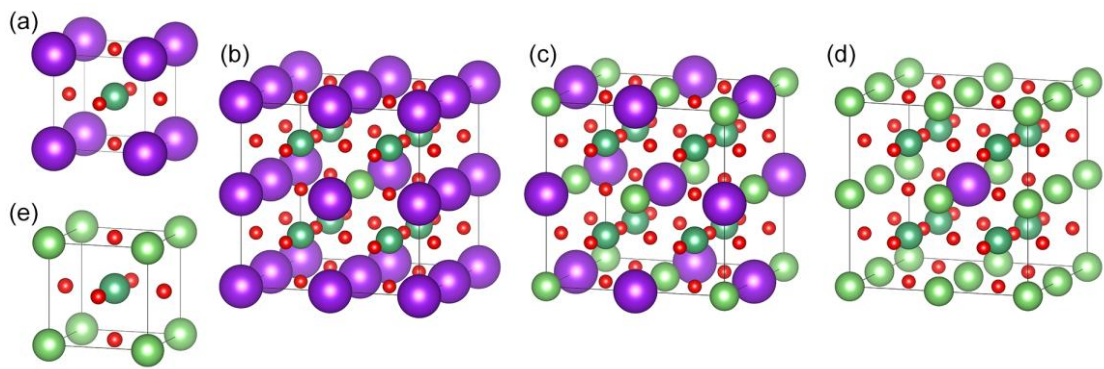

Figure S6. Atomic configurations of (a) KNO, (b) KLN125, (c) KLN500, (d) KLN875, (e) LNO. Purple, red, dark-green, and light-green atoms correspond to K, O, Nb, and Li, respectively.

Table S7. Estimation of the Shannon's ionic radius of Li.

| Coordination number | Existing (Å) | Extrapolated (Å)* |
|---|---|---|
| 4 | 0.59 | N/A |
| 6 | 0.76 | N/A |
| 9 | 0.92 | N/A |
| 12 | N/A | 1.13 (0.17) |

* The extrapolation is achieved based on the linear fitting of existing data. The fitted function is y = 0.345 + 0.065*x, where y and x represent the ionic radius and c coordination number, respectively. The standard errors for y and x are 0.0577 and 0.0087, respectively. The adjusted $R^2$ of the fitting is 0.965.

Table S8. Lattice parameters and volume of normalized tetragonal unit cell of different compositions.

| Parameters | KNO | KLN125 | KLN500 | KLN875 | LNO |
|---|---|---|---|---|---|
| $c$ (Å) | 4.205 | 4.234 | 4.278 | 4.201 | 4.184 |
| $a$ (Å) | 3.998 | 3.981 | 3.948 | 3.930 | 3.921 |
| $c/a$ | 1.052 | 1.063 | 1.084 | 1.069 | 1.067 |
| Vol. (Å$^3$) | 67.221 | 67.102 | 66.680 | 64.884 | 64.326 |